\pdfoutput=1
\documentclass{article}
\usepackage{hyperref}

\usepackage{authblk}

\usepackage{silence}
\usepackage{afterpage}

\usepackage{lipsum}% just to generate text for the example
\usepackage{fancyhdr} 
\usepackage{supertabular, setspace}
\usepackage[a4paper, total={7in,9in}]{geometry}
\usepackage[utf8]{inputenc}
\usepackage[title]{appendix}
\usepackage{graphics}\usepackage[demo]{graphicx}
\usepackage{amssymb,amsmath,amsthm,enumitem}

\usepackage{wrapfig}
\usepackage{subcaption}
\usepackage{amsmath}
\usepackage{mathtools}
\usepackage{amsfonts}
\usepackage{amssymb}
\usepackage{amsthm}
\newtheorem{theorem}{Theorem}[section]

\newtheorem{example}[theorem]{Example}

\newtheorem{remark}{\sc Remark}
\newtheorem{lemma}{\sc Lemma}[section]
\newtheorem{corollary}{\sc Corollary}[section]
\newtheorem{definition}{\sc Definition}[section]

\newcommand{\be}{\begin{eqnarray}}
	\newcommand{\ee}{\end{eqnarray}}
\newcommand{\Be}{\begin{eqnarray*}}
	\newcommand{\Ee}{\end{eqnarray*}}
\newcommand{\bee}{\begin{equation}}
	\newcommand{\eee}{\end{equation}}
\newcommand{\ba}{\begin{array}}
	\newcommand{\ea}{\end{array}}
\newcommand{\bl}{\begin{lemma}}
	\newcommand{\el}{\end{lemma}}
\newcommand{\bd}{\begin{definition}}
	\newcommand{\ed}{\end{definition}}
\newcommand{\bt}{\begin{theorem}}
	\newcommand{\et}{\end{theorem}}
\newcommand{\bp}{\begin{proof}}
	\newcommand{\ep}{\end{proof}}
\newcommand{\bi}{\begin{itemize}}
	\newcommand{\ei}{\end{itemize}}
\newcommand{\br}{\begin{remark}}
	\newcommand{\er}{\end{remark}}
\newcommand{\bc}{\begin{corollary}}
	\newcommand{\ec}{\end{corollary}}
\newcommand{\bex}{\begin{example}}
	\newcommand{\eex}{\end{example}}
\usepackage{chngcntr}

\usepackage{breqn} %by me
\usepackage{cite} % it put - btw citation and automatically arranges in order
\newlength{\bibitemsep}
\newlength{\bibparskip}
\let\oldthebibliography\thebibliography
\renewcommand\thebibliography[1]{%
  \oldthebibliography{#1}%
  \setlength{\parskip}{\bibitemsep}%
  \setlength{\itemsep}{\bibparskip}%
}
\allowdisplaybreaks

\newgeometry{margin=1in, bottom=.80in, top=.80in}
\counterwithin*{equation}{section}
\counterwithin*{equation}{section}
\begin{document}
\date{}

\title{On exact solutions, conservation laws and invariant analysis of the generalized  Rosenau-Hyman equation}

\author[1]{Pinki Kumari\thanks{yadav.pk1403@gmail.com}}
\author[1,2]{R.K. Gupta\thanks{rajeshateli@gmail.com}}
\author[1]{Sachin Kumar\thanks{Corresponding Author:\,sachin1jan@yahoo.com,\,sachin.kumar@cup.edu.in}}

\affil[1]{Department of Mathematics and Statistics,
		Central University of Punjab, Bathinda, Punjab, India}
\affil[2]{Department of Mathematics, Central University of Haryana, Mahendergarh, Haryana, India}

\maketitle

\begin{abstract}
In this paper, the nonlinear Rosenau-Hyman equation with time dependent variable coefficients is considered for investigating its invariant properties, exact solutions and conservation laws. Using Lie classical method, we derive symmetries admitted by considered equation. Symmetry reductions are performed for each components of optimal set. Also nonclassical approach is employed on considered equation to find some additional supplementary symmetries and corresponding symmetry reductions are performed.  Later three kinds of exact solutions of considered equation are presented graphically for different parameters. In addition, local conservation laws are constructed for considered equation by multiplier approach.\\
\textbf{Keywords:} Classical and Nonclassical symmetries; Rosenau Hyman equation; Conservation laws; Exact solution\\
\textbf{Mathematics Subject Classification:} 70S10;
76M60; 83C15
\end{abstract}

%\textbf{Mathematics Subject Classification:} 76M60, 35L65, 35R11, 35C10, 35C08.\\
%\textbf{Keywords:} Lie symmetry, Symmetry reduction, Power series solution, Conservation laws.
\section{Introduction}\label{sec1}
From past few decades, the theory of nonlinear differential equations has undergone notable achievements. Nonlinear partial differential equations are the mathematical formulations of laws of nature. Generally they appear in the mathematical analysis of diverse physical phenomena in the area of engineering, applied sciences and mathematical physics \cite{deb}. Most of the problems in physics are nonlinear and often hard to solve in explicit manner, so each nonlinear problem is studied as an individual problem. These nonlinear phenomena have several physical and mathematical applications which are usually interpreted by finding their numerical or analytic solutions. In order to obtain numerical solution, numerical, asymptotic and perturbation methods are usually used with great success; nonetheless, much interest prevails in the direction of finding closed form analytical solutions. Various effective analytical techniques \cite{khan,jafari,abbas,fan,feng,fu} have been developed to find exact solution in literature. Among these techniques, symmetry reduction techniques \cite{bluman,bluman1,clarkson,gandarias} are the most effective and straightforward, more generally, it is the general theory to construct exact solution in terms of solutions of lower dimensional equations and also the most active field of research nowadays.

A well known symmetry approach- Lie classical approach, originally proposed by Norwegian mathematician Sophus Lie, is completely algorithmic as it does not involve any kind of guesses and have gained popularity in obtaining symmetries, similarity transformations and symmetry reductions. Later nonclassical method is developed to find supplementary symmetries, which is not revealed by classical method. As nonclassical method involves nonlinear determining equations, which are complicated to solve. So this method is not exploited much. In this work, we will use both classical and nonclassical approach to explore symmetries and symmetry reduction for Rosenau-Hyman (RH) equation with time dependent variable coefficients. The RH equation with variable coefficients is written as
\begin{eqnarray}\label{1}
u_t+\alpha (t)uu_{xxx}+\delta(t)u_xu_{xx}+\beta(t)uu_{x}=0
\end{eqnarray}
where $u=u(x,t)$, the coefficients $\alpha(t),~\beta(t),~\delta(t)$ are nonzero integrable functions of $t$ and subscripts stand for partial derivatives. For $\alpha=\beta=1$ and $\delta=3$, this equation illustrate the formation of patterns in liquid drops \cite{ros}.

The assumption of constant coefficients usually leads to idealization of the physical phenomena in which nonlinear models appear. That is why, the study of nonlinear models with variable coefficients is gaining much attention nowadays. To analyse the sensitivity of physical situations with various significant parameters construed by variable coefficients, exact solutions of these models are very important and helpful. 

One of the popular aspect in the current study on nonlinear partial differential equations is the conservation laws. Mathematically conservation laws are the differential equation that describes natural laws. Though some conservation laws do not have a physical relevance, but in the context of partial differential equations, they reveals certain qualities like integrability, existence and uniqueness of solutions \cite{lax,benjamin,knops}. Besides, exact solutions of partial differential equation can be derived with the help of conserved vectors linked with Lie symmetries \cite{sjo,bok,cara}. First of all, Noether theorem \cite{noether} has been introduced to construct conservation laws for variational problems. Now several methods like direct construction method, new conservation method etc. \cite{anco1,anco2,kara,steudel} have been developed in which no priori knowledge about Lagrangian is needed.

The paper is arranged in the manner- In sec. \ref{sec:2}, Lie point symmetries of eq. \eqref{1} are obtained and symmetry reductions are performed. Sec. \ref{sec:3} deals with the nonclassical symmtries of \eqref{1}. In sec. \ref{sec:4}, exact solutions of \eqref{1} are found. Conservation laws of \eqref{1} are constructed in sec. \ref{sec:5}. Conclusion ends the paper in sec. \ref{sec:6}.

\section{Point Symmetries for Eq. (\ref{1})}\label{sec:2}
This section deals with the application of Lie classical method \cite{kumar,kawahara} to eq. (\ref{1}). The point symmetries obtained by this method allow reduction of PDE to ordinary differential equation. 

First we assume that a continuous group of point transformations $(x^{\oplus},t^{\oplus},u^{\oplus})$  in one parameter as
\begin{equation}
\begin{aligned}\label{2}
t^{\oplus} &=t+\epsilon\tau(x,t,u)+O(\epsilon^{2})\\
x^{\oplus} &=x+\epsilon\xi(x,t,u)+O(\epsilon^{2})\\
u^{\oplus} & =u+\epsilon\eta (x,t,u)+O(\epsilon^{2})
\end{aligned}
\end{equation}
that leaves (\ref{1}) invariant. Here, $\epsilon$ is a continuous group parameter and $\xi,\tau,\eta,\eta^t,\eta^x,\eta^{xx}, \eta^{xxx}$ are infinitesimals corresponding to $x,t,u,u_t,u_x,u_{xx},u_{xxx}$ respectively and extended infinitesimal are computed by the following formulas
\begin{equation}
\begin{aligned}\label{2.1}
&\eta^t=D_t(\eta)-u_xD_t(\xi)-u_tD_t(\tau)\\
&\eta^x=D_x(\eta)-u_xD_x(\xi)-u_tD_x(\tau)\\
&\eta^{xx}=D_x(\eta^x)-u_{xx}D_{xx}(\xi)-u_{tx}D_{xt}(\tau)\\
&\eta^{xxx}=D_x(\eta^{xx})-u_{xxx}D_{xxx}(\xi)-u_{txx}D_{xxt}(\tau)
\end{aligned}
\end{equation}
Also, we assume that the infinitesimal generator takes the form 
\begin{equation}\label{3}
X\equiv\xi(x,t,u)\frac{\partial}{\partial x}+\tau(x,t,u)\frac{\partial}{\partial t}+\eta(x,t,u)\frac{\partial}{\partial u}
\end{equation}
Invariance of transformation (\ref{2}) on (\ref{1}) leads the following invariance criterion
\begin{equation}\label{4}
\begin{aligned}
\eta ^t+\alpha(t)[\eta u_{xxx}+\alpha'(t)uu_{xxx}\tau +u\eta^{xxx}]+\beta(t)[\eta u_x+u\eta^x]+\beta '(t)uu_x\tau+\\\delta(t)[\eta^xu_xx+u_x\eta^{xx}]+\delta '(t)u_xu_{xx}=0
\end{aligned}
\end{equation}
Putting the values of extended infinitesimals (\ref{2.1}) in (\ref{4}) and comparing the coefficients of various linearly independent monomials, the following essential system of determining systems is obtained.
\begin{equation}
\begin{aligned}\label{5}
&\xi_u=\tau_x=\tau_u=0\\
&\alpha^2u^2\eta_{uuu}+\delta \alpha u \eta_{uu}=0\\
&3\alpha u\xi_x-\alpha_tu\tau-\alpha \eta-\tau_t \alpha u=0\\
&\alpha \beta u^2\eta_x+\alpha^2u^2\eta_{xxx}+\alpha u\eta_t=0\\
&3\alpha^2u^2\eta_{ux}-3\alpha^2u^2\xi_{xx}+\alpha \delta u\eta_x=0\\
&3\alpha^2u^2\eta_{uux}+2\delta \alpha u\eta_{ux}-\delta \alpha u\xi_{xx}=0\\
&3\alpha^2u^2\eta_{uu}-\alpha \delta \eta+\alpha \delta_t u\tau-\delta \alpha_t u\tau+\delta \alpha u\eta_u=0\\
&2\alpha \beta \xi_x u^2-\alpha\xi_tu-\alpha^2\xi_{xxx}u^2+3\eta_{xxu}\alpha^2u^2+\\&\delta\alpha u\eta_{xx}+\alpha \beta_t u^2\tau-\beta \alpha_tu^2\tau=0
\end{aligned}
\end{equation}
The general solution of (\ref{5}) is written as
\begin{equation}
\begin{aligned}\label{5.1}
&\tau=\frac{\int -c_1 \alpha(t)dt+c_2}{\alpha(t)}\\
&\xi= c_3x+c_4\\
&\eta= (c_1+3c_3)u\\
&\alpha=\alpha(t)\\
&\beta=c_5\Bigg(\Big(\int -c_1 \alpha(t)dt+c_2\Big)^{\frac{c_3}{c_1}}\Bigg)^2\alpha(t)\\
&\delta=c_6\alpha(t)
\end{aligned} 
\end{equation}
where $c_1,c_2~,c_3,~c_4,~c_5~\text{and}~c_6$ are arbitrary constants. If we take $c_3=0$, symmetries obtained in expression (\ref{5.1}) coincide with the symmetries of RH equation with coefficients $\alpha=\beta=1$ and $\delta=3$. So it is clear that symmetries (\ref{5.1}) are generalized version. Now we consider $c_3=c_1$ to ease our reduction calculation and for the case, symmetries take the form 
\begin{equation}
\begin{aligned}\label{5.2}
&\tau=\frac{-c_1\int \alpha(t)dt+c_2}{\alpha(t)},~~\xi= c_1x+c_4,~~\eta= 4c_1u\\
&\alpha=\alpha(t)\\
&\beta=c_5\Big(\int -c_1 \alpha(t)dt+c_2\Big)^2\alpha(t)\\
&\delta=c_6\alpha(t)
\end{aligned} 
\end{equation}
 The associated Lie algebra spanned by the infinitesimal generators (\ref{5.2}) is written as
 \begin{equation*}
 \begin{aligned}
 V_1=x\frac{\partial}{\partial x}-\frac{\int \alpha (t)dt}{\alpha(t)}\frac{\partial}{\partial t}+4u\frac{\partial}{\partial
  u}
  \end{aligned} 
   \end{equation*}
  \begin{equation}
   \begin{aligned}\label{5.21}
   V_2=\frac{\partial}{\partial x},~V_3=\frac{1}{\alpha(t)}\frac{\partial}{\partial t}
 \end{aligned} 
 \end{equation}
 
 We can perform symmetry reduction for any linear combination of aforementioned generators as any linear combination of generators again gives a generator. There are large number of such combinations. So to find non equivalent symmetry reductions, we use concept of optimal system \cite{olver}. Here, we find optimal set of vector fields with the components
 $$V_1~\text{and}~V_2+\lambda V_3$$ 
 Now we tabulate invariants, similarity variables, similarity transformation and variable coefficients for each component of optimal system in Table 1.\\
\begin{table}\label{tab:1}
 % title of Table
\centering % used for centering table
\caption{Similarity ansatz}
\begin{tabular}{c c c c} % centered columns (4 columns)
\hline\hline %inserts double horizontal lines
Generators & Invariants & Ansatz & Variable coefficients \\ [0.5ex] % inserts table
%heading
\hline % inserts single horizontal line
 &  &  &  $\alpha(t)=\alpha(t)$\\ % inserting body of the table
$V_1$ & $(x\int\alpha(t)dt,(\int\alpha(t)dt)^{4}u)$ & $u=(\int\alpha(t)dt)^{-4}f(x\int\alpha(t)dt)$ & $\beta(t)=-c_5(\int  \alpha(t)dt)^2\alpha(t)$ \\
 &  &  & $\delta(t)=c_6\alpha(t)$ \\
 &  &  & $\alpha(t)=\alpha(t)$ \\
$V_2+\lambda V_3$ & $(x-\frac{1}{\lambda}\int \alpha (t)dt,u)$ & $u=f(x-\frac{1}{\lambda}\int \alpha (t)dt)$ & $\beta(t)=c_5\lambda^2\alpha(t)$ \\
&  &  & $\delta(t)=c_6\alpha(t)$ \\ [1ex] % [1ex] adds vertical space
\hline %inserts single line
\end{tabular}
%\label{table:nonlin} % is used to refer this table in the text
\end{table}
\begin{theorem}
The similarity variable $\zeta=x\int\alpha(t)dt$ and similarity transformation $u=(\int\alpha(t)dt)^{-4}f(\zeta)$  for vector field $V_1$ reduces eq. (\ref{1}) into the following nonlinear ordinary differential equation.
\begin{equation}\label{5.3}
-4f+zf'+ff{'''}-c_5ff'+c_6f'f{''}=0
\end{equation}
Here, $'$ denotes the first order derivative w.r.t. $\zeta$.
\end{theorem}
\begin{theorem}
For the vector field $V_2+\lambda V_3$, the similarity transformation $u=f(\zeta)$ with similarity variable $\zeta=x-\frac{1}{\lambda}\int \alpha (t)dt$ reduces eq. (\ref{1}) into the following ODE
\begin{equation}\label{5.4}
-\frac{1}{\lambda}f'+ff{'''}+c_5\lambda^2ff'+c_6f'f{''}=0
\end{equation}
\end{theorem}

\section{Nonclassical Symmetries}\label{sec:3}
This section presents nonclassical symmetries of eq. \eqref{1} via compatibility conditions explained in Refs. \cite{yan,yan2}. First we consider nonclassical symmetry of the form
\begin{equation}\label{5.5}
u_t=\xi(x,t,u)u_x+\eta(x,t,u)
\end{equation}
In order to find compatibility condition, substitute \eqref{5.5} into \eqref{1} and expression reads as
 \begin{equation}\label{5.6}
 \xi u_x+\eta+\alpha (t)uu_{xxx}+\beta(t)uu_{x}+\delta(t)u_xu_{xx}=0
 \end{equation} 
 Next, equality of $u_{tt}$ between \eqref{1} and \eqref{5.5} yields the following equation
 \begin{dmath}\label{5.7}
  \xi_tu_x+\xi_uu_tu_x+\xi u_{xt}+\eta_t+\eta_uu_t+\alpha_tuu_{xxx}+\alpha u_tu_{xxx}+\alpha uu_{xxxt}+\beta_tuu_{x}+\beta u_tu_{x}+\beta uu_{xt}+\delta_tu_xu_{xx}\delta u_{xt}u_{xx}+\delta u_x u_{xxt}=0
  \end{dmath}
 Further, by eliminating differential consequences of $u_t$ and highest order term $u_{xxx}$ from eq. \eqref{5.7} with \eqref{5.5}  and \eqref{5.6}, we obtain the following set of nonlinear determining equations
\begin{equation}\label{5.8}
\begin{aligned}
u_{xx}:~~&\delta \eta_x-3\alpha u\xi_{xx}+3\alpha u\eta_{xu}=0\\
0:~~&\eta_{t}+\beta u\eta_x-\eta \frac{\alpha_t}{\alpha}+\alpha u\eta_{xxx}+3\eta\xi_x-\frac{\eta^2}{u}=0\\
u_xu_{xx}:~~&\delta_t-9\alpha u\xi_{xu}+3\alpha \eta_{uu}u-\frac{\eta\delta}{u}+\eta_u\delta-\delta\frac{\alpha_t}{\alpha}=0\\
u_x:~~&-\beta u\frac{\alpha_t}{\alpha}+\beta_tu+3\eta\xi_u-3\xi\xi_x+\delta\eta_{xx}-\xi_tu+\\~&\xi\frac{\alpha_t}{\alpha}+\frac{\eta\xi}{u}+2u\beta\xi_x+3\alpha\eta_{xxu}-\alpha u\xi_{xxx}=0\\
u_x^2u_{xx}:~~&-6\alpha u\xi_{uu}=0\\
u_x^2:~~&-3\xi\xi_{u}+2\delta\eta_{xu}-\delta\xi_{xx}+3u\beta\xi_u+3\alpha u\eta_{xuu}\\~&-3\alpha u\xi_{xxx}=0\\
u_x^3:~~&\delta\eta_{uu}-2\delta\xi_{xu}+\alpha u\eta_{uuu}-3\alpha u\xi_{xuu}=0\\
u_{xx}^2:~~&-3\alpha u\xi_u=0\\
u_x^4:~~&-\delta\xi_{uu}-\alpha u\xi_{uuu}=0
\end{aligned}
\end{equation}
and the solution of system \eqref{5.8} yields $\xi,~\eta$ and variable coefficients as
\begin{equation}
\begin{aligned}\label{5.9}
&\xi=f(t)(x+c_1)\\
&\eta=c_2f(t)u\\
&\alpha(t)=c_3f(t)\exp(-\int(c_2-3)f(t)dt)\\
&\beta(t)=c_4f(t)\exp(-\int(c_2-1)f(t)dt)\\
&\delta(t)=c_5f(t)\exp(-\int(c_2-3)f(t)dt)\\
\end{aligned}
\end{equation}
Here, $c_1,c_2,c_3,c_4,c_5$ are arbitrary constants and $f(t)$ is arbitrary integrable function.
Thus, nonclassical symmetries of variable coefficients RH equation \eqref{1} is written as
\begin{equation}
\begin{aligned}\label{5.10}
u_t+f(t)(x+c_1)u_x-c_2f(t)u\equiv0
\end{aligned}
\end{equation}
and the corresponding generator of symmetry is
\begin{equation}\label{5.101}
X\equiv f(t)(x+c_1)\frac{\partial}{\partial x}+\frac{\partial}{\partial t}+c_2uf(t)\frac{\partial}{\partial u}
\end{equation}
Now we will obtained similarity transformation for eq. \eqref{1} by solving the following characteristic equation
\begin{equation}\label{5.102}
\frac{dx}{f(t)(x+c_1)}=\frac{dt}{1}=\frac{du}{c_2f(t)u}
\end{equation}
For the nonclassical symmetries \eqref{5.101}, eq. \eqref{5.102} produces similarity variable as $z=(x+c_1)exp(-\int f(t)dt)$ and similarity transformation as $u=exp(c_2\int f(t)dt)g(z)$, which reduce the eq. \eqref{1} into the following ODE
\begin{equation}\label{5.103}
-zg'+c_2g+c_3gg{'''}+c_4gg'+c_5g'g{''}=0
\end{equation}

\section{Exact Solution}\label{sec:4}
In this section, we aim to find the exact solutions of eq. (\ref{1}) by seeking the solutions for reduced equations (\ref{5.3}), \eqref{5.4} and \eqref{5.103}. Also graphical representation will be shown by taking several different values of parameters. 
\subsection{Solution of Eq. \eqref{5.3}}
 Consider solution of eq. \eqref{5.3} has series form as 
\begin{equation}\label{6}
 \begin{aligned}
f(\zeta)=a+b\zeta+c\zeta^2
 \end{aligned}
\end{equation}
where $a,~b,~\text{and}~c$ are unknown constants that need to be determined.\\
On putting (\ref{6}) into (\ref{5.3}), we obtain
\begin{equation}\label{7}
 \begin{aligned}
 a=0,~~b=-\frac{3}{c_5},~~c=0
 \end{aligned}
\end{equation}
So by reverting back the original variables, the exact solution of \eqref{1} can be written as
\begin{equation}
\begin{aligned}\label{8}
u(x,t)=-\frac{3}{c_5}x\Big(\int\alpha(t)dt\Big)^{-3}
\end{aligned}
\end{equation}
3D and contour plots of the solution \eqref{8} is shown in figures (a)-(i) with some different parameters.\\
\begin{figure}[htb]\label{fig1}
  \centering
      \includegraphics[width=0.30\linewidth]{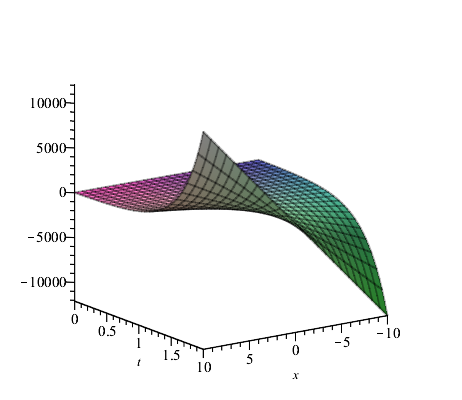}\hfill
      \includegraphics[width=0.25\linewidth]{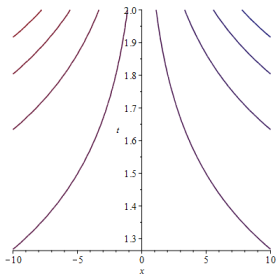}\hfill
       \includegraphics[width=0.30\linewidth]{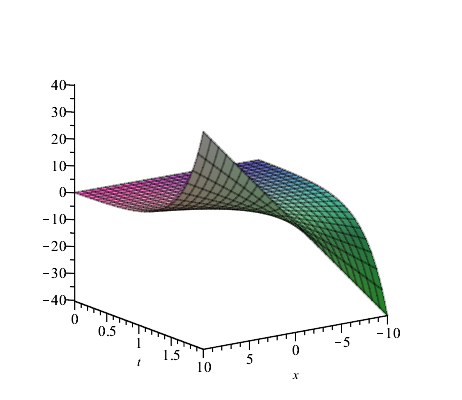}\\
       (a) $\alpha(t)=e^{-t}$, $c_5=1$\hfill
       (b) $\alpha(t)=e^{-t}$, $c_5=1$\hfill
       (c) $\alpha(t)=e^{-t}$, $c_5=300$\\
      \includegraphics[width=0.30\linewidth]{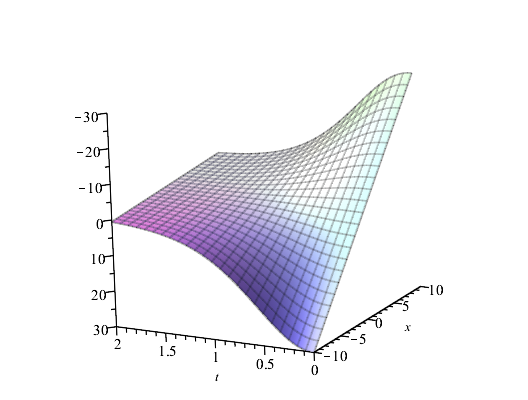}\hfill
      \includegraphics[width=0.30\linewidth]{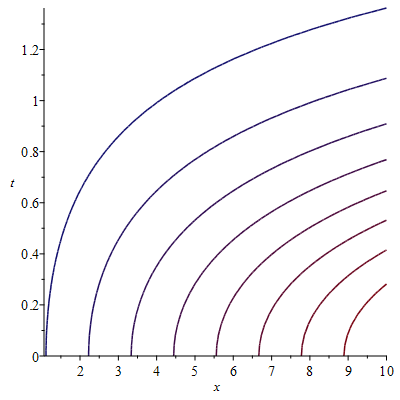}\hfill
       \includegraphics[width=0.30\linewidth]{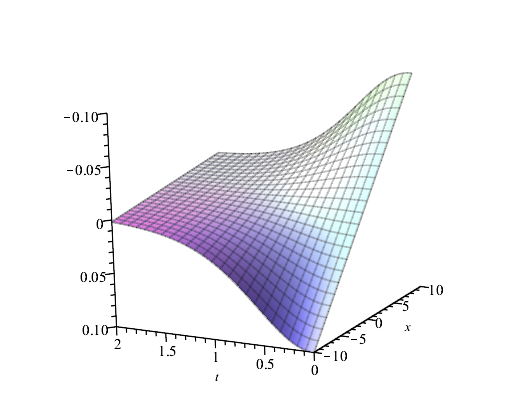}\\
       (d) $\alpha(t)=sinh(t)$, $c_5=1$\hfill
       (e) $\alpha(t)=sinh(t)$, $c_5=1$\hfill
       (f) $\alpha(t)=sinh(t)$, $c_5=300$\\
        \includegraphics[width=0.30\linewidth]{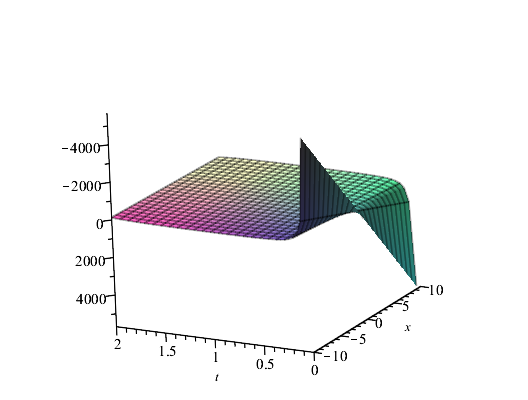}\hfill
       \includegraphics[width=0.30\linewidth]{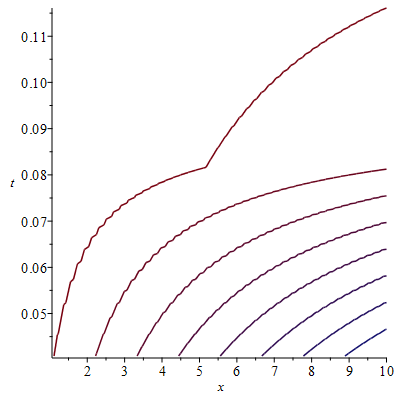}\hfill
       \includegraphics[width=0.30\linewidth]{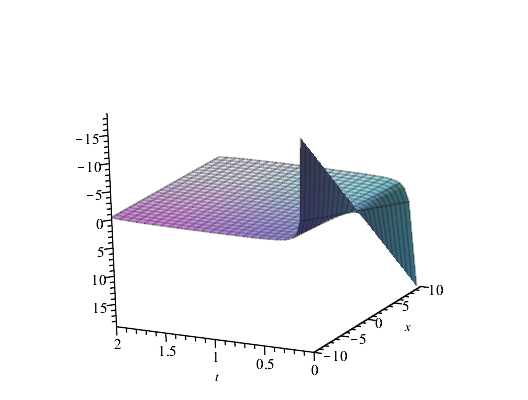}\\
       (g) $\alpha(t)=ln(t)$, $c_5=1$\hfill
       (h) $\alpha(t)=ln(t)$, $c_5=1$\hfill
       (i) $\alpha(t)=ln(t)$, $c_5=300$\\
  %\caption{caption here}
\end{figure}
\subsection{Solution of Eq. \eqref{5.4}}
For arbitrary constant $c_6$, the work to find exact solution is in progress. Presently, we seek the solution for $c_6=1$. Now eq. \eqref{5.4} is written as
\begin{equation}
\begin{aligned}\label{9}
-\frac{1}{\lambda}f'+ff{'''}+c_5\lambda^2ff'+f'f{''}=0
\end{aligned}
\end{equation}
Eq. \eqref{9}, on integrating  w.r.t. $\zeta$, can be expressed as   
\begin{equation}
\begin{aligned}\label{10}
-\frac{1}{\lambda}f+ff{''}+\frac{c_5\lambda^2}{2}f^2=0
\end{aligned}
\end{equation}
With the aid of Maple software, the general solution of eq. \eqref{10} is obtained as
\begin{equation}
\begin{aligned}\label{11}
f(z)=c_1sin(\frac{1}{\sqrt{2}\lambda}\sqrt{c_5}z)+c_2cos(\frac{1}{\sqrt{2}\lambda}\sqrt{c_5}z)+\frac{2}{c_5\lambda^3}
\end{aligned}
\end{equation}
Consequently, the general solution admitted by \eqref{1} can be written as
\begin{dmath}\label{12}
%\begin{aligned}
u(x,t)=c_1sin\Big(\frac{1}{\sqrt{2}}\lambda\sqrt{c_5}(x-\frac{1}{\lambda}\int \alpha (t)dt)\Big)+c_2cos\Big(\frac{1}{\sqrt{2}}\lambda\sqrt{c_5}(x-\frac{1}{\lambda}\int \alpha (t)dt)\Big)+\frac{2}{c_5\lambda^3}
%\end{aligned}
\end{dmath}
3D profile of solution \eqref{12} is displayed in figures (j)-(o) and contour plots are shown in figures (j.1)-(l.1) with some parametric values. 
\begin{figure}[htb]
  \centering
      \includegraphics[width=0.30\linewidth]{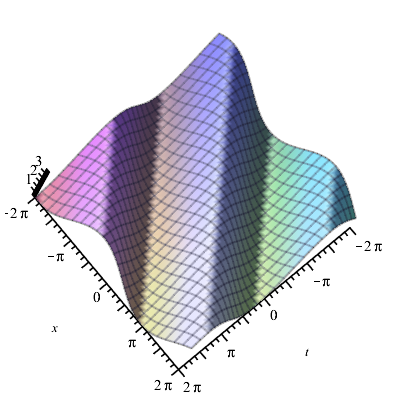}\hfill
      \includegraphics[width=0.30\linewidth]{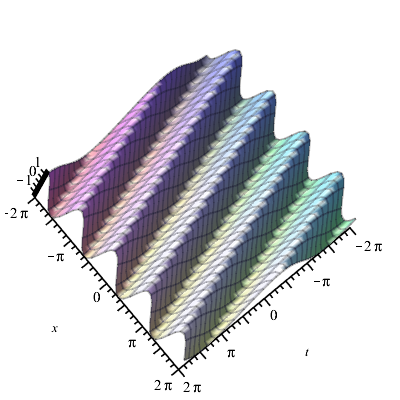}\hfill
       \includegraphics[width=0.30\linewidth]{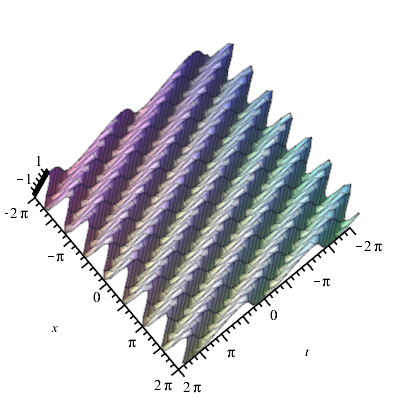}\\
       (j) $\alpha$(t)=1, $c_5$=1, $c_1$=1,\hfill
       (k) $\alpha$(t)=1, $c_5$=1, $c_1$=1,\hfill       
       (l) $\alpha$(t)=1, $c_5$=3, $c_1$=1, \\
        $~~~~c_2$=1, $\lambda$=1\hfill
              $c_2$=1, $\lambda$=3\hfill       
               $c_2$=1, $\lambda$=3 \\
      \includegraphics[width=0.25\linewidth]{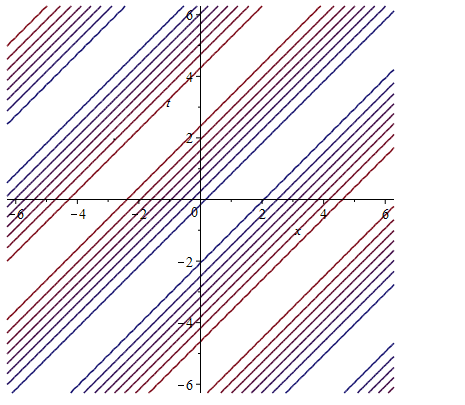}\hfill
      \includegraphics[width=0.25\linewidth]{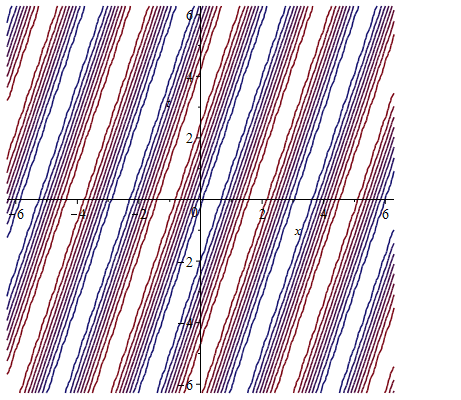}\hfill
       \includegraphics[width=0.25\linewidth]{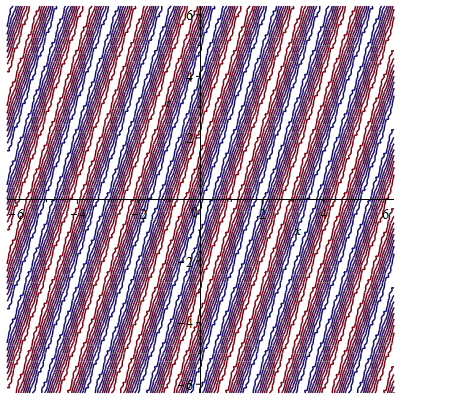}\\
       (j.1) $\alpha$(t)=1, $c_5$=1, $c_1$=1,\hfill
       (k.1) $\alpha$(t)=1, $c_5$=1, $c_1$=1, \hfill 
       (l.1) $\alpha$(t)=1, $c_5$=3, $c_1$=1, \\
        $~~~~c_2$=1, $\lambda$=1\hfill
         $c_2$=1, $\lambda$=3\hfill
        $c_2$=1, $\lambda$=3\\
        \includegraphics[width=0.30\linewidth]{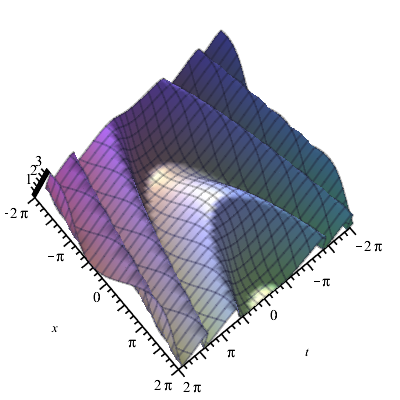}\hfill
       \includegraphics[width=0.30\linewidth]{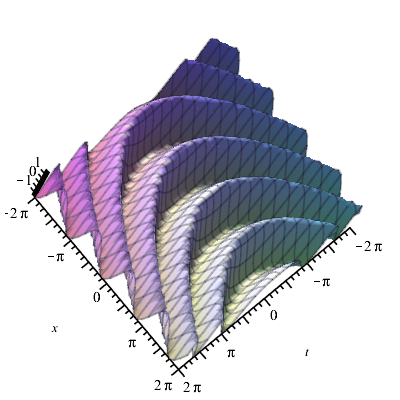}\hfill
       \includegraphics[width=0.30\linewidth]{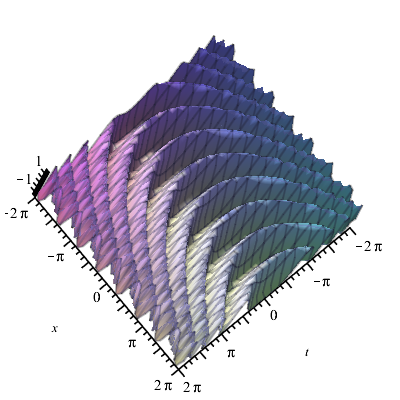}\\
       (m) $\alpha$(t)=t, $c_5$=1, $c_1$=1,\hfill
       (n) $\alpha$(t)=t, $c_5$=1, $c_1$=1,\hfill
       (o) $\alpha$(t)=t, $c_5$=3, $c_1$=1,\\
        $~~~~c_2$=1, $\lambda$=1\hfill
         $c_2$=1, $\lambda$=3\hfill
          $c_2$=1, $\lambda$=3\\
       \end{figure}
\subsection{Solution of Eq. \eqref{5.103}}
We assume that the eq. \eqref{5.103} takes the series solution as follows 
\begin{equation}\label{5.104}
 \begin{aligned}
g(z)=a+bz+cz^2
 \end{aligned}
\end{equation}
where $a,~b,~\text{and}~c$ are unknown constants need to be determined.
By substituting (\ref{5.104}) into (\ref{5.103}), we get
\begin{equation}\label{7.1}
 \begin{aligned}
 a=0,~~b=\frac{1-c_2}{c_4},~~c=0
 \end{aligned}
\end{equation}
Thus by reverting back the original variables, the exact solution of \eqref{1} is found as
\begin{equation}
\begin{aligned}\label{8.1}
u(x,t)=\frac{1-c_2}{c_4}(x+c_1)\exp\Big(\int(c_2-1)f(t)dt\Big)
\end{aligned}
\end{equation}
3D graph and contour plots of \eqref{8.1} are represented in figures (p)-(u) by appointing various function parameters.
\begin{figure}[htb]
  \centering
      \includegraphics[width=0.30\linewidth]{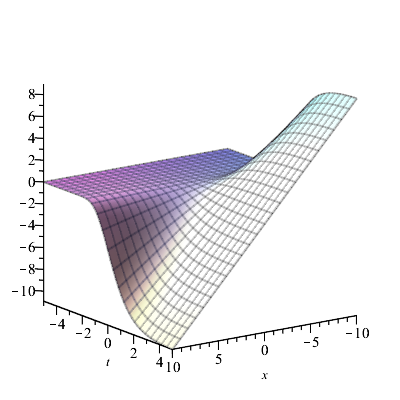}\hfill
      \includegraphics[width=0.30\linewidth]{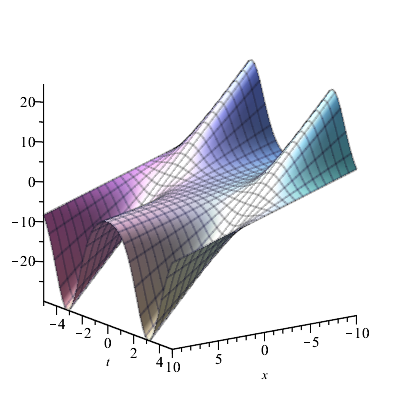}\hfill
       \includegraphics[width=0.30\linewidth]{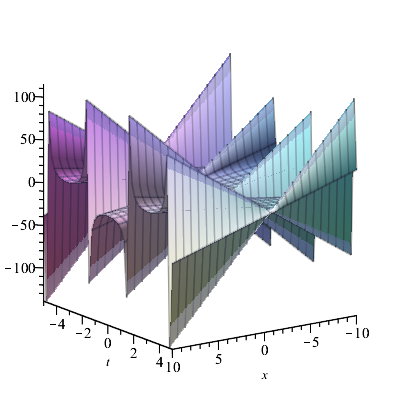}\\
       (p) $f(t)=exp(-t)$, $c_1$=1, \hfill
       (q)$f(t)=sin(t)$, $c_1$=1, \hfill       
       (r)$f(t)=tan(t)$, $c_1$=1, \\
       $~~~~c_2$=2, $c_4$=1\hfill
       $c_2$=2, $c_4$=1\hfill
       $c_2$=2, $c_4$=1\\
       \includegraphics[width=0.30\linewidth]{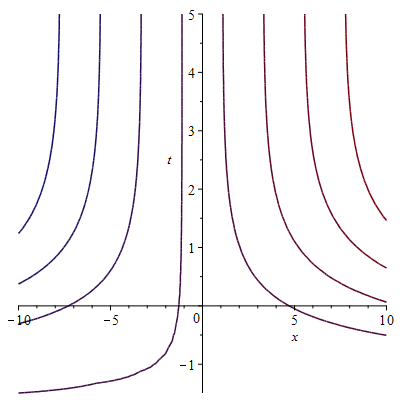}\hfill
             \includegraphics[width=0.30\linewidth]{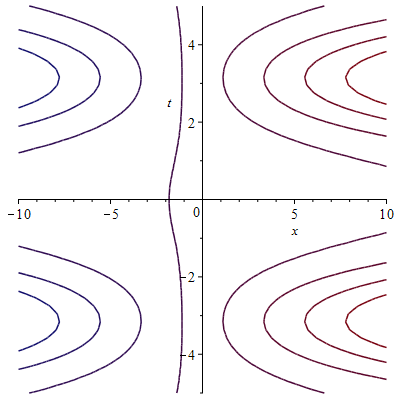}\hfill
              \includegraphics[width=0.30\linewidth]{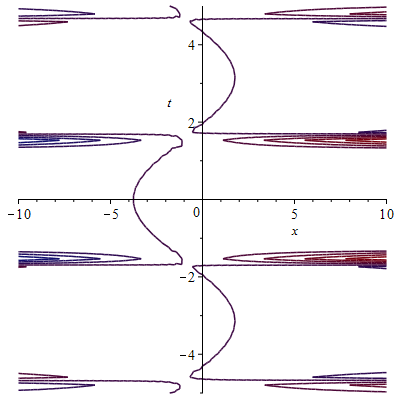}\\
              (s) $f(t)=exp(-t)$, $c_1$=1, \hfill
              (t)$f(t)=sin(t)$, $c_1$=1, \hfill       
              (u)$f(t)=tan(t)$, $c_1$=1, \\
              $~~~~c_2$=2, $c_4$=1\hfill
              $c_2$=2, $c_4$=1\hfill
              $c_2$=2, $c_4$=1\\
\end{figure}

\section{Conservation Laws}\label{sec:5}

Here we intend to construct local conservation laws of RH equation \eqref{1} by multiplier method \cite{naz,steudel}. In this method, no priori knowledge about Lagrangian of equation of motion is necessary. The method is quite well known, effective and it directly uses the definition of local conservation laws.     
First we assume simple multiplier of the form $\Lambda(x,t,u)$. The multiplier $\Lambda$ for (\ref{1}) have the property
\begin{equation}\label{19.1}
\Lambda (u_t+\alpha (t)uu_{xxx}+\beta(t)uu_{x}+\delta(t)u_xu_{xx})=D_tC^t+D_xC^x
\end{equation}
for all solutions of $u(x,t)$. Here, $D_t~\text{and}~D_x$ are total derivative operator defined by 
\begin{equation}
\begin{aligned}\label{19.2}
D_t=&\frac{\partial}{\partial t}+u_t\frac{\partial}{\partial u}+u_{tt}\frac{\partial}{\partial u_t}+u_{xt}\frac{\partial}{\partial u_x}+\cdots\\
D_x=&\frac{\partial}{\partial x}+u_x\frac{\partial}{\partial u}+u_{tx}\frac{\partial}{\partial u_t}+u_{xx}\frac{\partial}{\partial u_x}+\cdots
\end{aligned}
\end{equation}
 The determining equation for the multiplier $\Lambda$ follows by 
\begin{dmath}\label{20}
%\begin{aligned}
\frac{\delta}{\delta u}\Big[\Lambda (u_t+\alpha (t)uu_{xxx}+\beta(t)uu_{x}+\delta(t)u_xu_{xx}=0\Big]_{\eqref{1}}=0
%\end{aligned}
\end{dmath}
where Euler operator, $\frac{\delta}{\delta u}$ is expressed as
%\begin{equation}
\begin{dmath}\label{20.1}
\frac{\delta}{\delta u}=\frac{\partial}{\partial u}-D_t\frac{\partial}{\partial u_t}-D_{x}\frac{\partial}{\partial u_x}+D_{xx}\frac{\partial}{\partial u_{xx}}+D_{tt}\frac{\partial}{\partial u_{tt}}+D_{xt}\frac{\partial}{\partial u_{xt}}+\cdots
\end{dmath}
%\end{equation}
On expansion of \eqref{20}, the following set of determining equations is obtained.
\begin{equation}\label{21}
\begin{aligned}
&\Lambda_{x}\delta-3\Lambda_x\alpha-3\Lambda_{ux}\alpha u =0\\&2\Lambda_{ux}\delta-6\Lambda_{ux}\alpha-3\Lambda_{uux}\alpha u=0\\&
\Lambda_{xx}\delta-3\Lambda_{xx}\alpha-3\Lambda_{xux}\alpha u=0\\&\Lambda_{uu}\delta-3\Lambda_{uu}\alpha-\Lambda_{uuu}\alpha u=0,\\&
-\Lambda_{x}\beta-3\Lambda_{xxx}\alpha u-\Lambda_{t}=0\\&3\Lambda_{u}\delta-6\Lambda_u\alpha-3\Lambda_{uu}\alpha u=0
\end{aligned}
\end{equation}
Now we split the general solutions of eq. \eqref{21} into cases with different pivots in order to get maximum number of conserved vectors. The conserved vectors are constructed for each case explicitly.\\
\textbf{Case (a)} If $\alpha\neq0,~\delta(\delta-3\alpha)\neq0,~\delta_t\alpha-\delta\alpha_t\neq0$, 
the multiplier takes the form
\begin{equation}\label{22}
\Lambda=c_1
\end{equation}

 For this case, the conserved components of \eqref{1} are written as
\begin{equation}
\begin{aligned}\label{23}
&C^t=c_1u\\
&C^x=\alpha c_1uu_{xx}-\frac{1}{2}\alpha c_1u_x^2+\frac{1}{2}\delta c_1u_x^2+\frac{1}{2}\beta c_1u^2
\end{aligned}
\end{equation}
\textbf{Case (b)} For $\alpha\neq0,~\delta(\delta-3\alpha)\neq0,~\delta\neq0,~\delta_t\alpha-\delta\alpha_t=0$, two subcases arise.

\textbf{Subcase (b.1)}For $\delta=c_3\alpha,~c_3(c_3-3)\neq0$, the multiplier is
\begin{equation}
\begin{aligned}\label{24}
\Lambda=c_1+c_2u^{c_3-1}
\end{aligned}
\end{equation}

 So the conserved vectors associated with multiplier \eqref{24} are expressed as
\begin{equation}
\begin{aligned}\label{25}
C^t=&\frac{c_2}{c_3}u^{c_3}+c_1u\\
C^x=&\alpha c_1uu_{xx}+\alpha c_2uu_{xx}+\frac{1}{2}\beta c_1u^2+\frac{1}{2}\alpha c_1 c_3u_x^2-\frac{1}{2}c_1\alpha c_1u_x^2+\frac{\beta c_2}{c_3+1}u^{c_3+1}
\end{aligned}
\end{equation}

\textbf{Subcase (b.2)} The multiplier for $\delta=\alpha$ takes the form
\begin{equation}\label{26}
\Lambda=c_2ln(u)+c_1
\end{equation}

The expression for conserved vectors with \eqref{26} are as follows
\begin{equation}
\begin{aligned}\label{27}
C^t=&u(c_2ln(u)+c_1-c_2)\\
C^x=&\alpha c_2ln(u)uu_{xx}+\alpha c_1uu_{xx}-\frac{1}{2}\alpha c_2u^2_x+\beta c_2\Big(\frac{1}{2}u^2ln(u)-\frac{1}{4}u^2\Big)+\frac{1}{2}c_1u^2
\end{aligned}
\end{equation}
\textbf{Case (c)} For $\alpha\neq0,~\delta(\delta-3\alpha)=0,~\delta\neq0,~\delta_t\beta-\delta\beta_t\neq0$, we obtain the following multiplier
\begin{equation}
\begin{aligned}\label{28}
\Lambda=c_2u^2+c_3
\end{aligned}
\end{equation}

 So the conserved fluxes for multiplier \eqref{28} are expressed as
\begin{equation}
\begin{aligned}\label{29}
C^t=&\frac{1}{3}c_2u^3+c_3u\\
C^x=&\alpha c_2u^3u_{xx}+\frac{1}{4}\beta c_2u^4 +\alpha c_3u_x^2+\frac{1}{2}\beta c_3u^2+\alpha c_3uu_{xx}
\end{aligned}
\end{equation}
\textbf{Case (d)} For $\alpha\neq0,~\delta(\delta-3\alpha)=0,~\delta\neq0,~\delta_t\beta-\delta\beta_t=0$, two subcases are encountered.

\textbf{Subcase (d.1)} If $\alpha=\frac{1}{3}\delta,~\beta=\frac{1}{3}c_5\delta,c_5>0$, the multiplier is obtained as
\begin{equation}
\begin{aligned}\label{30}
\Lambda=c_1u^2+c_2+c_3sin(\sqrt{c_5}x)+c_4 cos(\sqrt{c_5}x)
\end{aligned}
\end{equation}

 The conserved vectors for multiplier \eqref{30} are computed as
\begin{equation}
\begin{aligned}\label{31}
C^t=&\frac{1}{3}c_1u^3+c_2u+c_3usin(\sqrt{c_5}x)+c_4 cos(\sqrt{c_5}x)u\\
C^x=&\frac{1}{12}\delta(4c_2u_x^2+c_1c_5u^4+2c_2c_5u^2+4c_3u_x^2sin(\sqrt{c_5}x) +4c_4u_x^2cos(\sqrt{c_5}x))+\\&4c_1u^3u_{xx}+4c_2uu_{xx}+4c_4\sqrt{c_5}usin(\sqrt{c_5}x)u_x-4c_3\sqrt{c_5}ucos(\sqrt{c_5}x)u_x\\&+4c_4\sqrt{c_5}ucos(\sqrt{c_5}x)u_{xx}+4c_3\sqrt{c_5}usin(\sqrt{c_5}x)u_{xx}
\end{aligned}
\end{equation}

\textbf{Subcase (d.2)} If $\alpha=\frac{1}{3}\delta,~\beta=-\frac{1}{3}c_5\delta,c_5>0$, the multiplier is
\begin{equation}
\begin{aligned}\label{32}
\Lambda=c_1u^2+c_2+c_3e^{\sqrt{c_5}x}+c_4 e^{-\sqrt{c_5}x}
\end{aligned}
\end{equation}

The conserved vectors for multiplier \eqref{32} are expressed as
\begin{equation}
\begin{aligned}\label{33}
C^t=&\frac{1}{3}c_1u^3+c_2u+c_3ue^{\sqrt{c_5}x}+c_4 e^{-\sqrt{c_5}x}u\\
C^x=&\frac{1}{12}\delta \Big(4c_2u_x^2-c_1c_5u^4-2c_2c_5u^2+4c_3u_x^2e^{\sqrt{c_5}x} +4c_4u_x^2e^{-\sqrt{c_5}x}+4c_1u^3u_{xx}+\\&4c_2uu_{xx}+4c_4\sqrt{c_5}ue^{-\sqrt{c_5}x}u_x-4c_3\sqrt{c_5}ue^{\sqrt{c_5}x}u_x+\\&4c_4\sqrt{c_5}ue^{-\sqrt{c_5}x}u_{xx}+4c_3\sqrt{c_5}ue^{\sqrt{c_5}x}u_{xx}\Big)
\end{aligned}
\end{equation}
\textit{Remark 1} It is verified that conservation laws of the original RH equation with $\alpha=\beta=1$ and $\delta=3$ are same as explained in sub-case (d.1).\\
\textit{Remark 2} In case (a), we found only one conserved vector. Two conserved vectors are encountered in case (b) and case (c). In both sub-cases of case (d), we get four conserved fluxes.

  \section{Conclusion}\label{sec:6}
  The work investigates the effectiveness of Lie classical method on Rosenau-Hyman equation with time dependent variable coefficients. Also, generalized supplementary symmetries, which was not revealed by classical theory, are obtained by nonclassical approach  and  symmetry reductions for both type of symmetries are performed successfully. Exact solutions of RH equation for each component of optimal set, generated by three vector fields, as well as for nonclassical symmetries are presented. The extracted solutions are discussed graphically by taking different values of parameters and may be significant in dynamic study of RH model. Moreover we have tried to find maximum number of conserved vectors for RH model \eqref{1} by multiplier approach. The obtained conserved vectors can be used to find the solution of eq. \eqref{1} and will be subject to future work.
  \section*{Acknowledgement}
  The autor, Pinki Kumari expresses her gratitude to the University Grants Commission for financial assistance (Grant no. 19/06/2016(i)EU-V).

\bibliographystyle{plain}
%\bibliography{interactnlmsample}

\end{document}